\documentclass[%
 twocolumn,
superscriptaddress,
preprintnumbers,
 amsmath,amssymb,
 prl,
]{revtex4-1}
\usepackage[dvipsnames]{xcolor}
\usepackage[pagebackref=false]{hyperref}
\hypersetup{
    colorlinks  =   true,
    linkcolor   =   BlueViolet,
    citecolor   =   BlueViolet,
    filecolor   =   BlueViolet,
    urlcolor    =   BlueViolet}
\usepackage{graphics,amssymb,amsmath,epsfig,color}
\usepackage{graphicx}
\usepackage{dcolumn}
\usepackage{bm}
\usepackage{color}
\usepackage{ulem}

\newcommand{\cmt}[1]{}

\begin{document}

\title{Radiatively-cooled quantum microwave amplifiers \cmt{operated at elevated temperatures}} 


\author{Mingrui Xu}
\altaffiliation[]{These authors contributed equally to this work.}
\affiliation{Department of Electrical Engineering, Yale University, New Haven, Connecticut 06520, USA}

\author{Yufeng Wu}
\altaffiliation[]{These authors contributed equally to this work.}
\affiliation{Department of Electrical Engineering, Yale University, New Haven, Connecticut 06520, USA}

\author{Wei Dai}
\affiliation{Department of Applied Physics, Yale University, New Haven, Connecticut 06520, USA}

\author{Hong X. Tang}
\email{hong.tang@yale.edu}
\affiliation{Department of Electrical Engineering, Yale University, New Haven, Connecticut 06520, USA}

\date{\today}

\begin{abstract}
Superconducting microwave amplifiers are essential for sensitive signal readout in superconducting quantum processors. Typically based on Josephson Junctions, these amplifiers require operation at milli-Kelvin temperatures to achieve quantum-limited performance. Here we demonstrate a quantum microwave amplifier that employs radiative cooling to operate at elevated temperatures. This kinetic-inductance-based parametric amplifier, patterned from a single layer of high-$T_\mathrm{c}$ NbN thin film\cmt{in the form of a nanobridge}, maintains a high gain and meanwhile enables low added noise of 1.3 quanta when operated at 1.5 Kelvin. Remarkably, this represents only a 0.2 quanta increase compared to the performance at a base temperature of 0.1 Kelvin. By uplifting the parametric amplifiers from the mixing chamber without compromising readout efficiency, this work represents an important step for realizing scalable microwave quantum technologies.
\end{abstract}

\maketitle
Superconducting parametric amplifiers are critical components for high-sensitivity readout of superconducting quantum processors \cite{aumentado2020}. To achieve fast, high-fidelity qubit readout \cite{Jeffrey2014fast, PhysRevApplied.7.054020}, the first-stage amplifier must operate with near quantum-limited noise performance \cite{caves1982}. Traditionally, Josephson Junction-based amplifiers operated at milli-Kelvin temperatures have been used for this purpose \cite{Yurke1989,yamamoto2008, castellanos-beltran2008, abdo2011josephson}. However, with the increasing number of superconducting qubits \cite{arute2019quantum, kim2023evidence}, the demand for an increasing number of readout lines poses significant challenges in terms of space occupation and power consumption at the mixing chamber (MXC) stage of dilution fridges \cite{meenehan2014silicon, meenehan2015pulsed, hornibrook2015cryogenic, midolo2018nano}. Despite the fact that the amplifier itself may have a small footprint, the magnetic shield, isolators, and circulators associated with it still occupy considerable space. Similarly, while the power required to drive the parametric amplifier may be insignificant, the attenuators necessary for reducing background noise contribute to significant power dissipation. These factors collectively pose challenges to the scalability of microwave quantum technologies.

An ideal superconducting parametric amplifier should maintain quantum-limited noise regardless of its material temperature. However, current state-of-the-art parametric amplifiers are typically limited to operation at milli-Kelvin temperatures due to several reasons. First, the superconducting transition temperature sets a hard ceiling to the operating temperature of the devices. Second, the intrinsic loss within the amplifier introduces excessive fluctuations to the signal by thermalizing the mode temperature to the material temperature of the device. The latter is further exacerbated by thermal quasiparticle population as $k_B T$ gets comparable to superconducting bandgap $\hbar \Delta$. Therefore, the use of Josephson-junction-based amplifiers is restricted to a temperature well below 1K, the critical temperature of aluminum from which they are fabricated. Hence, developing microwave parametric amplifiers using materials with high superconducting transition temperature and high intrinsic quality factor within the desired temperature range would be preferable. 

Kinetic inductance nonlinearity based on single layer NbN, NbTiN, and granular Al materials has drawn increasing attention in recent years \cite{vissers2015frequency, samkharadze2016high, maleeva2018circuit}. By utilizing a nanowire design, amplifiers predicated on kinetic inductance have demonstrated high gain with quantum-limited noise \cite{parker2022degenerate, xu2023-ct, khalifa2023nonlinearity}. Traveling-wave amplifiers have also been demonstrated with excellent performance compared with their Josephson junction counterparts \cite{ho2012wideband, chaudhuri2017broadband, esposito2021perspective, malnou2021three, malnou2022performance}.  Notably, these materials possess high critical temperatures, typically around 10 K for NbN and NbTiN \cite{cheng2019superconducting, sowa2017plasma, malnou2022performance}. This characteristic facilitates the operation of the kinetic-inductance traveling-wave parametric amplifier (KI-TWPA) at 4 K \cite{malnou2022performance}. As a result, amplifiers that incorporate these materials can sustain superconductivity even at higher temperatures.

In this paper, we introduce the nanobridge kinetic-inductance parametric amplifier (NKPA) \cite{xu2023-ct} that leverages the radiative-cooling concept \cite{XuPRL2020, Wang2021-hd, Albanese2020-kv} to achieve ultralow-added-noise microwave signal amplification. We are the first to demonstrate that, with radiative cooling in effect, the NKPA achieves near quantum-limited amplification performance with added noise of $n_\mathrm{add}=1.33\pm0.04$ at an operating temperature of 1.5~K.
These findings not only solidify our understanding of added noise versus physical temperature of parametric amplifiers, but also introduce a new operating regime for superconducting parametric amplifiers. This advancement paves the way for enhanced scalability in superconducting quantum computing and many sensing applications. 

\begin{figure}[h]
\includegraphics{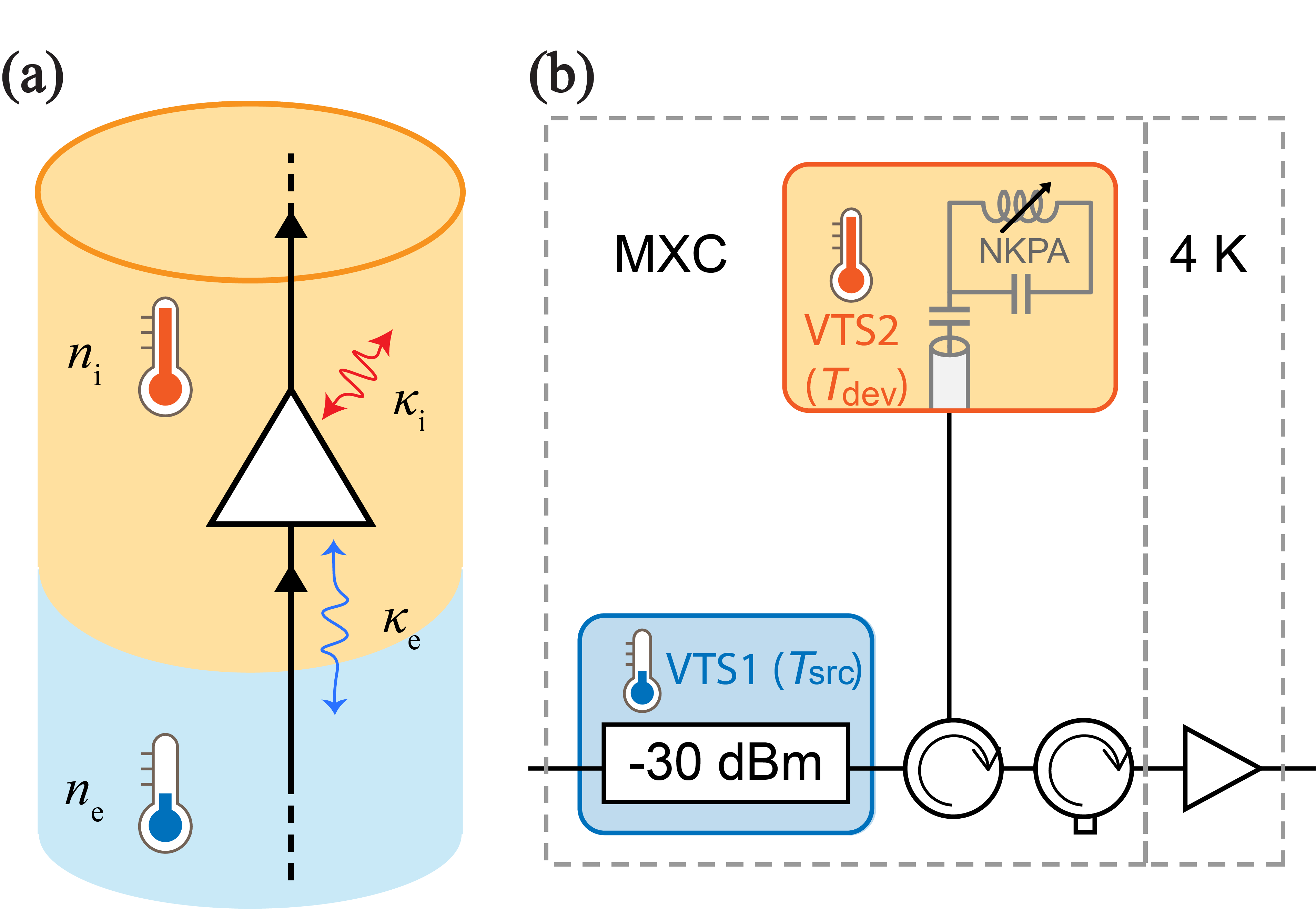}
\caption{\label{diagram} (a) Principle of radiatively-cooled amplification. To achieve radiative cooling for the amplifier installed in a hot enviroment with thermal occupancy $n_\mathrm{i}$, the amplifier is connected to a cooling bath through external coupling channel with thermal occupancy $n_\mathrm{e}$. The decay rate of the amplifier resonance through intrinsic and external coupling channels are denoted as $\kappa_\mathrm{i}$ and $\kappa_\mathrm{e}$, respectively.
(b) Schematic of the experimental setup. The cooling bath at temperature $T_\mathrm{src}$ is implemented as a 30 dB attenuator (near perfect 50 ohm load) anchored on a variable temperature stage (VTS1). The amplifier is mounted on a second variable temperature stage (VTS2) that defines the device material temperature $T_\mathrm{dev}$.}
\end{figure}

In the context of parametric amplification, radiative cooling could help reduce the added noise of a hot amplifier device ($T_{\rm{dev}}$) by using a cooling channel connected to a cold thermal bath ($T_{\rm{src}}$), as illustrated in Fig. \ref{diagram} (a). The thermal occupancy of the cavity mode $n_{\mathrm{mode}}$ is related to the thermal occupancy of the physical bath $n_{\mathrm{i}}$ and cold source $n_{\mathrm{e}}$, each corresponding to $T_{\rm{dev}}$ and $T_{\rm{src}}$ through Bose–Einstein distribution. The mode occupancy could be expressed as \cite{XuPRL2020}
\begin{equation}
\label{equ:radiative_cooling}
    n_{\mathrm{mode}} = \frac{\kappa_{\mathrm{e}}}{\kappa} n_{\mathrm{e}} +  \frac{\kappa_{\mathrm{i}}}{\kappa} n_{\mathrm{i}} 
\end{equation}[floatfix]
where $\kappa_{\mathrm{e}}$ and $\kappa_{\mathrm{i}}$ are the external coupling rate and the internal loss rate respectively, and $\kappa = \kappa_{\mathrm{e}} + \kappa_{\mathrm{i}}$ is the total loss rate, and $n_\mathrm{i}$ and $n_\mathrm{e}$ are related to $T_{\rm{dev}}$ and $T_{\rm{src}}$ through Bose-Einstein distribution $n_{\mathrm{s}}(T) = 1/\left(\exp(\hbar \omega_s/k_B T) - 1 \right)$. In order to minimize the thermalization of the mode with its physical environment, our strategy is to maximize the ratio of the external coupling rate (to the cable connected to the cold source) over the internal loss rate (to the warmer physical bath). This radiative cooling scheme~\cite{XuPRL2020, Wang2021-hd, Albanese2020-kv} has been utilized to prepare a superconducting resonator near its quantum ground state in the presence of environmental thermal excitations.

\begin{figure}[t]
    \centering
    \includegraphics[width=\linewidth]{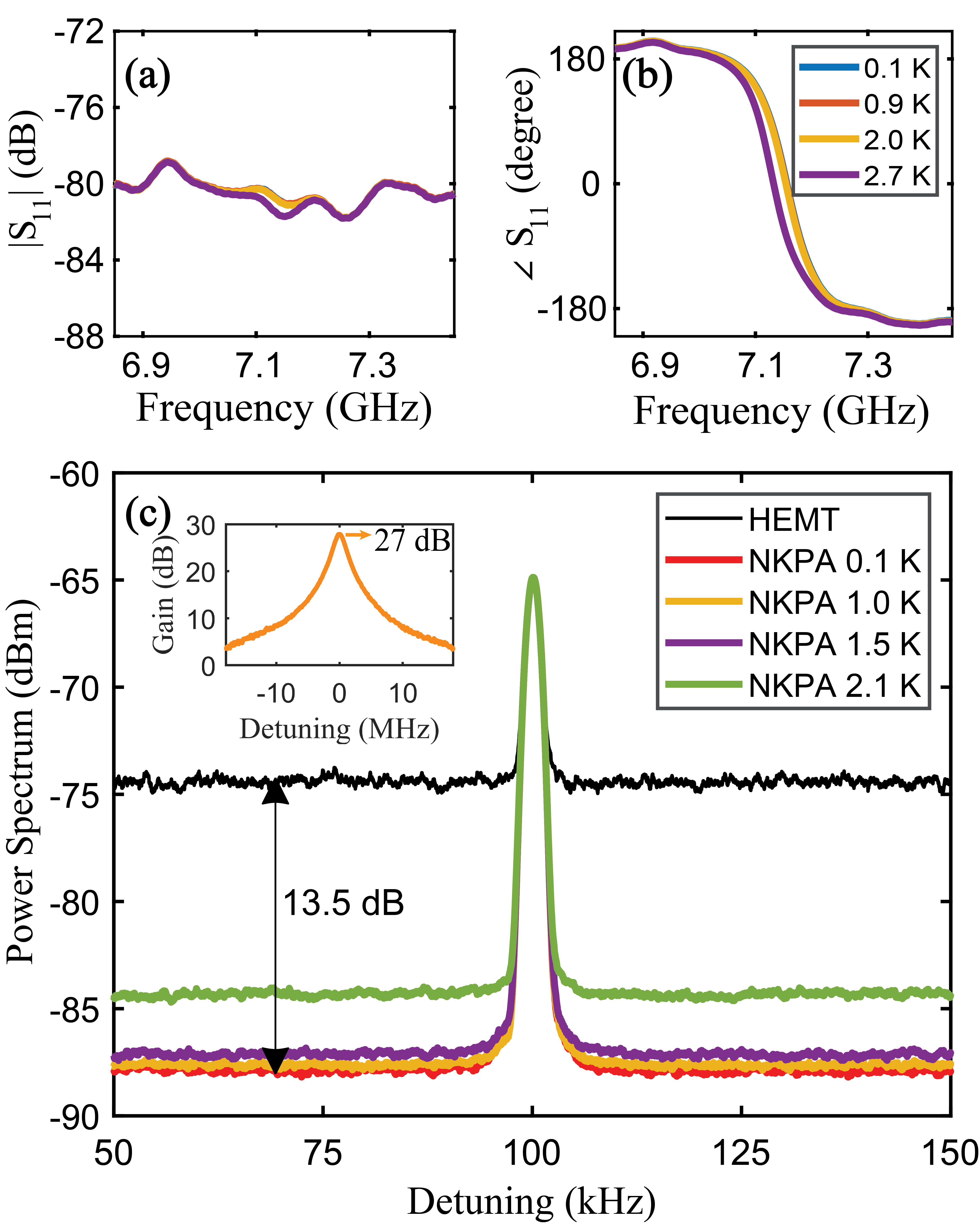}
               \caption{Device characterization and amplifier performance with varied device operating temperature. (a) The magnitude and (b) the phase response of the NKPA in the linear regime when the device material temperature increases from 0.1 K to 2.7 K. (c) Signal-to-noise ratio improvement with NKPA turned off (black line) and on (colored lines) in the phase-preserving mode with device operating temperature from 0.1 K to 2.1 K. The gain for the probe tone, detuned from the center amplification frequency by 100 kHz, is maintained at 27 dB for different temperatures. The power spectrum, measured with a resolution bandwidth of 1.8 kHz, is artificially offset to maintain a consistent power level for the output probe tone. The insert illustrates the gain profile while the NKPA was functioning as an amplifier. }
    \label{fig:gainspectrum}
\end{figure}

Our selected amplifier, the NKPA, is patterned from a high $T_\mathrm{c}$ NbN film on a silicon substrate. It has previously been demonstrated to be resilient to magnetic fields of up to 0.5 Tesla~\cite{xu2023-ct}. The device features a nanobridge with a cross-section of $80\, \text{nm} \times 4\, \text{nm}$, which provides a single photon Kerr nonlinearity of approximately $10\, \text{kHz}$. The resonance frequency of the device is $7.151\, \text{GHz}$. At base temperature of $0.13\, \text{K}$, the external coupling rate is found to be around $2\pi \times 65\, \text{MHz}$ from the reflection spectrum fitting \cite{mcrae2020materials}. However, because of the intentionally designed strong overcoupling, we could not extract the accurate internal loss rate from the reflection spectrum in the presence of background ripples in the frequency spectrum as shown in Fig.~1(a). Through radiative cooling performance reported later, we infer that the internal loss rate is below $2\pi \times 3$ MHz. 

This particular amplifier is mounted on a variable temperature stage (VTS2, see Fig.~\ref{diagram}) consisting of a heater and a calibrated thermometer. VTS2 is connected to the mixing chamber of a dilution refrigerator through a weak thermal link made of a stainless steel post, and allows the device operating temperature to vary from 130~mK to above 3~K while the rest of the mixing chamber maintains at 50~mK. Another variable temperature stage (VTS1), which serves as a reference thermal noise source \cite{xu2023-ct}, is implemented with a 30~dB attenuator mounted on it. The diagram of the setup is shown in Fig.~1(b). After the NKPA, we use a High Electron Mobility Transistor (HEMT) amplifier as a second-stage amplifier, with several stages of circulators in between. The output signal is then monitored by a vector network analyzer (VNA) and a spectrum analyzer (SA). 

By probing the linear reflection spectrum of NKPA while varying the device operating temperature, it is confirmed that the device maintains very overcoupled condition from milli-Kelvin temperatures up to 2.7~Kelvin, owing to the high critical temperature of NbN. No significant frequency shift is observed until 2~K, as shown in Fig. ~\ref{fig:gainspectrum}(a) and (b). We use a two-tone drive scheme to operate the amplifier, with each drive tone detuned by over 105~MHz from the NKPA's center frequency. Phase-preserving amplification up to 45~dB is recorded at up to 1.5 K as discussed in detail in Appendix. Upon turning on the amplifier in the phase-preserving amplification mode, we observed a 13.47~dB improvement in signal-to-noise ratio using a weak coherent signal as a reference, shown in Fig.~\ref{fig:gainspectrum}(c). As we increase the device's material temperature to 1.5 K, the noise floor increased only by 0.74 dB, indicating the amplifier performance was only marginally affected. The excellent noise performance is maintained up to 2.1 K, at which point the noise floor increased more dramatically by 3.55 dB.

\subsection{Added Noise Analysis}
\begin{figure}[t]
\includegraphics[width=\linewidth]{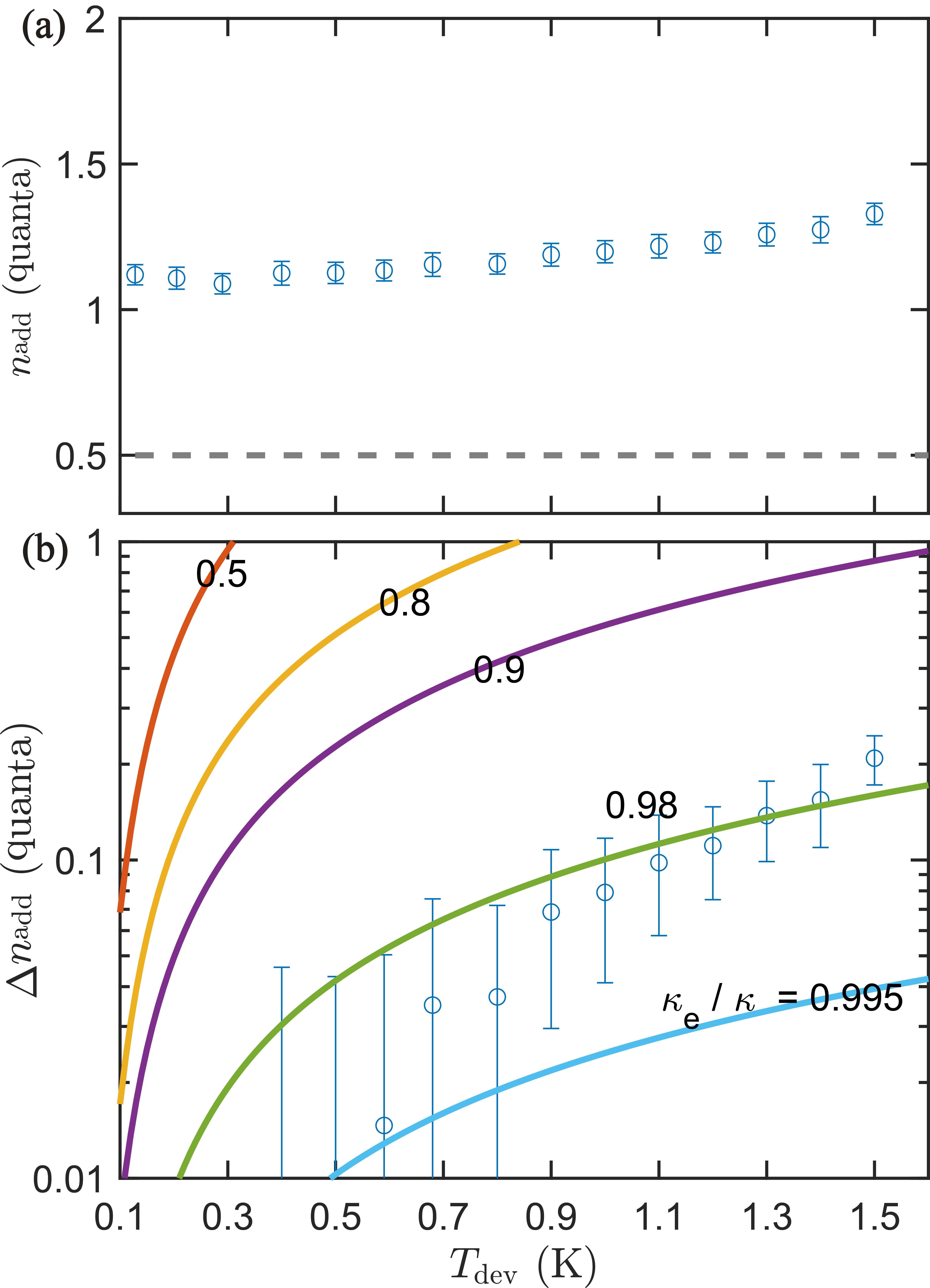}
\caption{\label{results} Added noise results from noise thermometry calibration as a function of the NKPA's operating temperature $T_\mathrm{dev}$. (a) The added noise in quanta referred to the output of the VTS, where the quantum-limited added noise of an ideal parametric amplifier is at 0.5. (b) The excessive added noise when NKPA is operated at elevated temperatures compared with the base temperature (0.13 K). The solid lines represent the expected excessive added noise level with different external coupling ratios $\kappa_{\mathrm{e}}/\kappa$. }
\end{figure}

To calibrate the amplifier added noise, we sweep the temperature of VTS1 to generate a reference thermal noise to feed to the NKPA \cite{malnou2018optimal, XuPRL2020}. As illustrated in Fig.\ref{results} (a), results show the total added noise of NKPA (referred to the input) is at $1.12\pm0.03$ quanta when operated at 130~mK, which is 0.62 quanta above (about twice of) the quantum limit. The excessive noise likely originates from unaccounted nonlinear processes in NKPA. Remarkably, the added noise of NKPA does not significantly increase while the operating temperature of the device increases to 1.5~K. Even at an operating temperature of 1 K, the added noise remains below 1.2 quanta.

To further investigate the influence of device operating temperature on the NKPA added noise, we attempt to understand the excessive added noise of NKPA as a function of the NKPA operating temperatures. Note that throughout this experiment, the temperature of the VTS1 remains at 130~mK, so the observed increase in added noise should be attributed to the resonator's thermalization to its material temperature through phonon dissipation. In the high-gain limit, the excessive noise due to increased device material temperature could be expressed as:
\begin{equation}
    \Delta n_\mathrm{add} = 2\frac{k_\mathrm{i}}{k_\mathrm{e}}n_\mathrm{dev} = 2\left( \frac{\kappa}{k_\mathrm{e}} - 1 \right) n_\mathrm{dev}
\end{equation}
\noindent where $\kappa=\kappa_e+\kappa_i$ represents the resonator linewidth and $n_\mathrm{dev}$ denotes the thermal photon occupation of device material temperature controlled by VTS2. The results and predicted curves based on various external coupling ratios $\kappa_\mathrm{e}/\kappa$ are shown in Fig.~\ref{results} (b). We found that the data fit relatively well to the predicted results with $\kappa_\mathrm{e}/\kappa=0.98$, indicating very overcoupled resonance and a decent radiative cooling effect. We can further infer the internal Q of the resonance of approximately 2000. In this case, we suspect the internal loss of NKPA is not likely limited by quasiparticles \cite{PhysRevB.72.014517,SurfaceImpedanceSuperconductors1991}.
    
When we further increase the device operating temperature beyond 2~K, the added noise increases drastically in a fashion that is not explainable by our linear model, which used both internal and external dissipation rates reflected in the reflective spectra of NKPA in the linear regime. Coincidentally, the increase of noise is accompanied by a pronounced shift in resonance frequency as the temperature increases, similar to the result reported by Gr{\"u}nhaupt et al. \cite{PhysRevLett.121.117001} from a thin-film granular Aluminum resonator.
Hence, we believe that spurious effects due to thermal quasiparticles come into play as temperature increases, which results in a drastic added noise increase to NKPA at around 2K.

\subsection{Prospective application}
\begin{figure}[h]
\includegraphics[width=\linewidth]{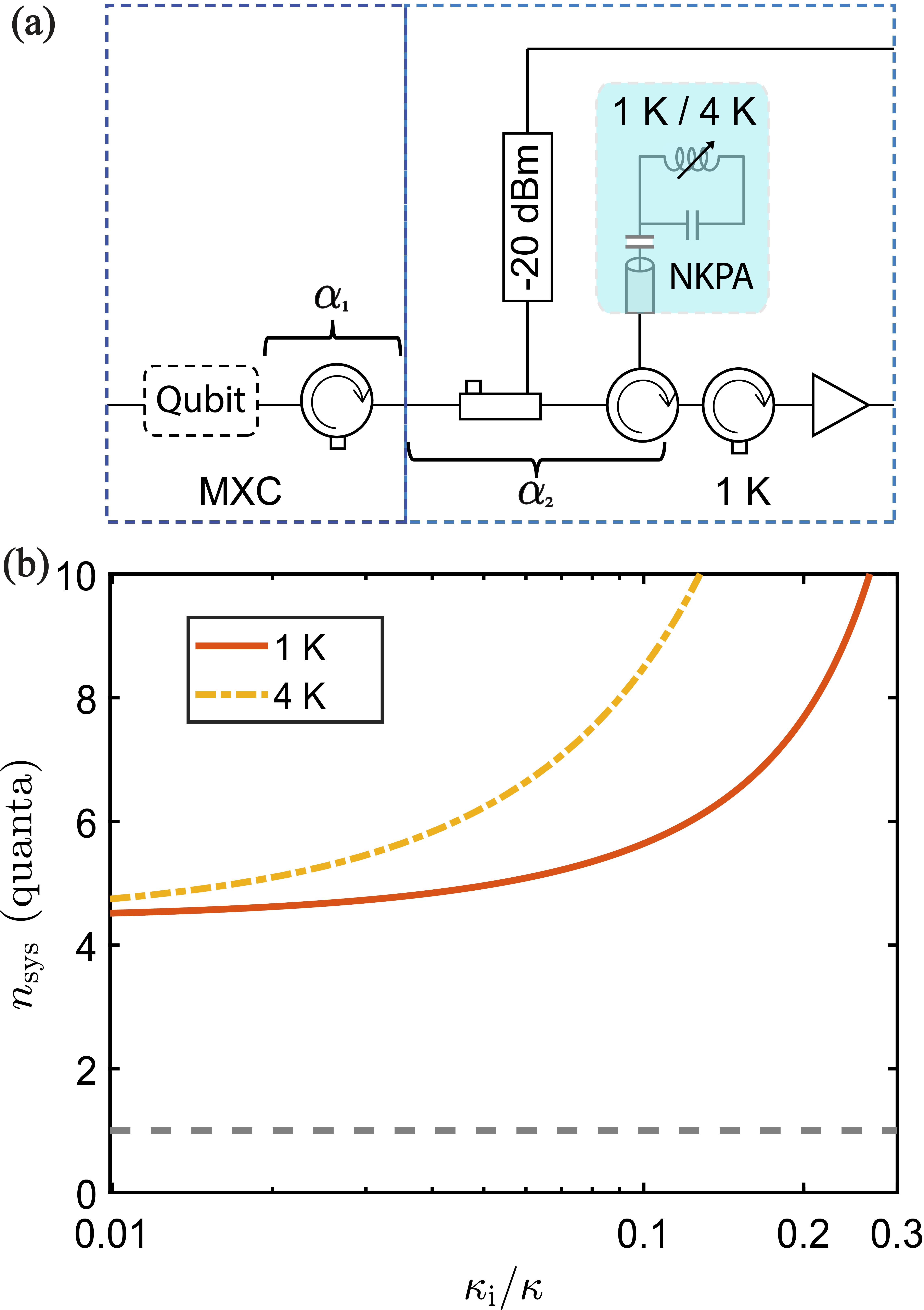}
\caption{\label{application_diagram}
The perspective of qubit readout using radiatively cooled NKPA. (a) Experimental scheme of qubit readout using parametric amplifier at 1-K/4-K Plate: the qubit is placed on the mixing chamber (MXC), followed by a circulator to shield against reflected pump power. The NKPA can be installed on either the 1-K or the bottom of the 4-K plate. A directional coupler is used to apply the drive power to the NKPA. (b) Theoretical estimation of readout line noise $n_\mathrm{sys}$ referred to the output of Qubit as a function of the internal coupling ratio $\kappa_{\mathrm{i}}/\kappa$ of radiatively cooled NKPA. The power transmission coefficient is assumed as $\alpha_1 = 0.9$ and $\alpha_2 = 0.8$. The gray dashed line indicates the quantum limit of the system noise ($n_\mathrm{sys} = 1$) with a phase-preserving parametric amplifier.
}
\end{figure}

The radiatively-cooled parametric amplifier demonstrated in this study could potentially address some of the scalability challenges faced by superconducting quantum computers, particularly the spatial and cooling power constraints of dilution refrigerators.
To better understand the perspective of using radiatively-cooled microwave amplifiers, in the following, we attempt to estimate the performance of a readout line incorporating radiatively cooled NKPA installed at 1-K or 4-K plates. 
We acknowledge that for the particular above-mentioned NKPA device discussed in this manuscript, the radiative cooling model only applies when the operating temperature is below 2 K. However, a NbTiN-based KI-TWPA is recently reported to exhibit satisfactory performance operating at a 4 K \cite{malnou2022performance}, suggesting that the issue of excessive added noise above 2 K for NKPA device discussed above could be addressed by improving the material choice and device engineering. 

A proposed configuration of utilizing radiatively cooled NKPA for quantum signal readout is depicted in Fig.~\ref{application_diagram}(a).
The primary benefit of shifting the NKPA and necessary circulators to the 1-K plate is that it liberates precious refrigeration resources within the mixing chamber. To feed drive power to the NKPA device, we propose to use a directional coupler thermalized at the 1-K plate, as opposed to mixing chamber for most junction-based ampliifers. Since the drive power does not need to be routed through the mixing chamber, its active heat load to the mixing chamber is thus estimated to be 0.1\,$\mu$W per device. For practical applications such as qubit readout, typically, an isolator is required after the qubit to prevent any backactions from the first-stage amplifier.  Due to the insertion loss of each component between the signal source and the amplifier, the signal-to-noise ratio of the entire output line will be inevitably degraded. Here, we denote the power transmission coefficient for the readout pulse at the mixing chamber and 1-K plate as $\alpha_1$ and $\alpha_2$. Thus, we can express the anticipated added noise of the system as follows:

\begin{equation}
\begin{split}
     n_{\mathrm{sys}} = & \frac{1}{\alpha_1 \alpha_2} \left[ \frac{\kappa_\mathrm{i}}{\kappa_\mathrm{e}} \left(2n_{\mathrm{s}}(T_{\mathrm{dev}}) +\frac{1}{2}\right) + n_{\mathrm{exc}} + 1\right] \\
     + & \frac{\kappa}{\kappa_e} \left[ 2\frac{1 - \alpha_1}{\alpha_1} \left( n_{\mathrm{s}}(0.01K) +1\right)  + 2\frac{1 - \alpha_2}{\alpha_1 \alpha_2} \left( n_{\mathrm{s}}(1K) + 1\right) \right],
\end{split}    
\end{equation}
where $n_\mathrm{exc}$ denotes the excessive noise of the amplifier. From this expression, we can evaluate the system noise impacted by the attenuation of each component. We assume a loss of 0.5~dB ($\alpha_1 \approx 0.9$) for the microwave output line in the mixing chamber, and a loss of 1~dB ($\alpha_2 \approx 0.8$) in the 1~K plate \cite{walter2017rapid, parker2022degenerate, simbierowicz2022microwave}. To give a realistic estimation, we retained the excessive noise value from the noise calibration, $n_\mathrm{exc}=0.62$. The simulated system noise referred to the output of the signal source (e.g. qubit readout pulse) is illustrated in Fig.~\ref{application_diagram}(b) with device operating temperatures of 1-K and 4-K. Our findings suggest that for the same level of ratiative cooling effectiveness demonstrated in this work, i.e. $\kappa_\mathrm{e}/\kappa = 0.98$ (or $\kappa_\mathrm{i}/\kappa = 0.02$),  the total output line noise is at 4.6 quanta when the device is at the 1-K plate. This value marginally increases to 5.1 quanta at the 4-K plate. This performance is competitive with that of the state-of-the-art Josephson Traveling-Wave Parametric Amplifiers (JTWPA) operating at the mixing chamber \cite{macklin2015near, white2015traveling}.

To maintain overcoupling at high temperatures, e.g. 4~K, we believe it would be helpful to implement NKPA made from superconducting films with higher $T_\mathrm{c}$ to mitigate the TLS and quasiparticle loss \cite{mcrae2020materials}. Materials such as NbTiN \cite{Yen1967-fb, malnou2022performance} or MBE-grown NbN \cite{Katzer2020-bs} can be explored. 

In conclusion, we demonstrate a radiatively-cooled superconducting parametric amplifier that achieves noise performance close to the quantum limit even at operating temperatures above 1~K. This advance is made possible by employing the kinetic-inductance nanobridge amplifier technology with high-$T_\mathrm{c}$ NbN films. Our results not only provide valuable insights into the impact of device material temperature on the excessive noise observed in parametric amplifiers but also hold tremendous potential for enabling rapid single-shot qubit readout for large-scale quantum computers by reducing heat load and space requirements within the mixing chamber. Ultimately, these findings contribute to enhancing the scalability of superconducting quantum computing devices. 

\subsection{Acknowledgments}
The authors would like to thank Professor Michel Devoret, Dr. Gangqiang Liu, Dr. Alessandro Miano, for useful discussions. We thank Dr. Yong Sun, Dr. Lauren McCabe, Mr. Kelly Woods, Dr. Michael Rooks, and Dr Sihao Wang for assistance in device fabrication. We acknowledge funding support from the Office of Naval Research on the development of nitride-based superconductors (under Grant No. N00014-20-1-2126)\cmt{and from
Army Research Office on the quantum transducer development (through Grant No. W911NF-18-1-0020)}. The part of the research that involves cryogenic instrumentation is supported by the DOE Office of Science, National Quantum Information Science Research Centers, Co-design Center for Quantum Advantage (C2QA), Contract No. DE-SC0012704.

\section{Appendix}
\subsection{Added-noise calibration}

\begin{figure}
    \centering
    \includegraphics[width=\linewidth]{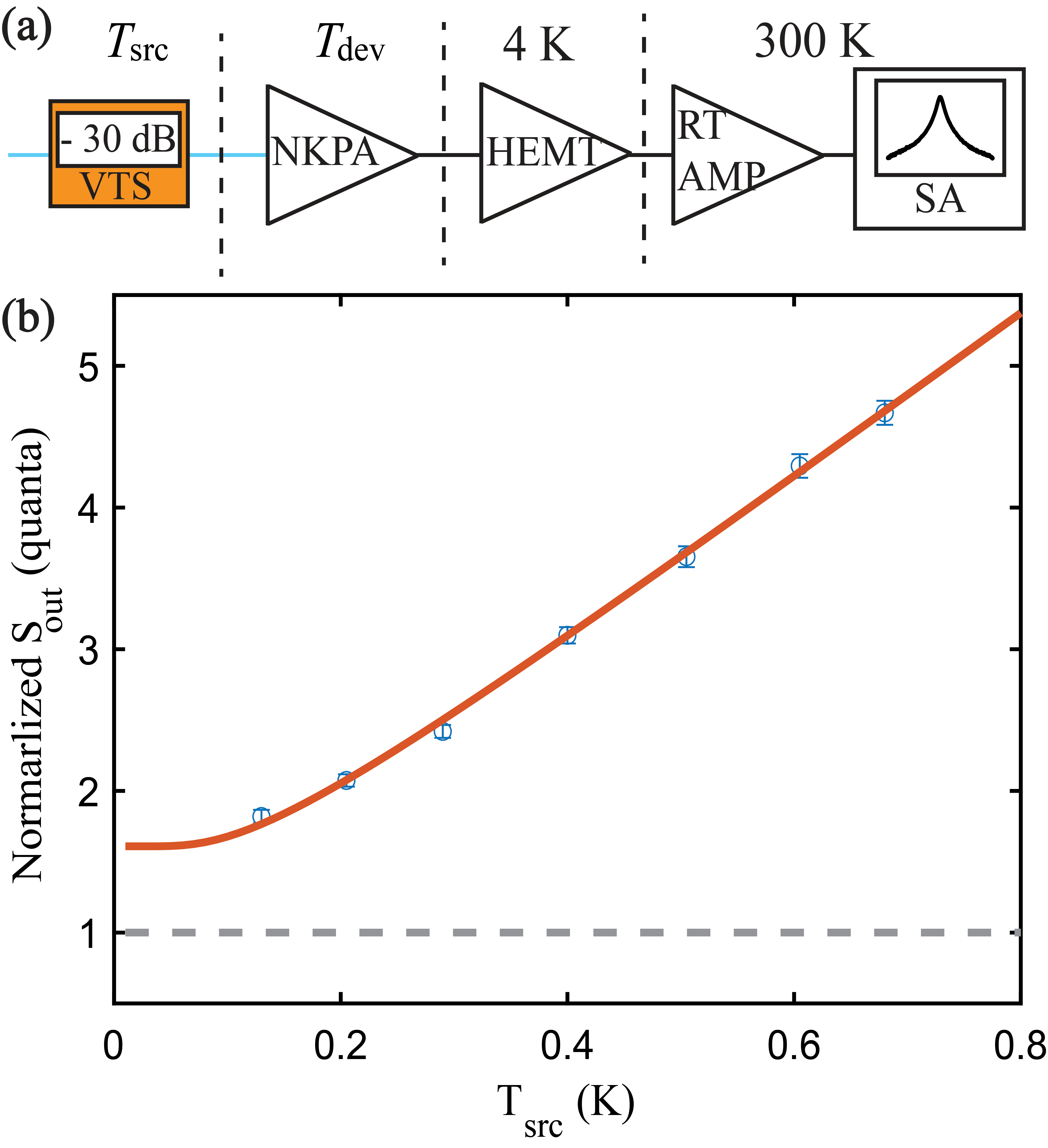}
    \caption{(a) Noise thermometry calibration setup diagram. The reference thermal noise sourced from the VTS is preamplified by the NKPA, which operates at a 27 dB gain, and then further amplified by a high-electron-mobility transistor (HEMT). The noise power spectrum is measured by a spectrum analyzer (SA). The cyan line represents superconducting cables. (b) Output noise power spectrum in quanta as a function of the thermal source temperature $T_\mathrm{src}$. }
    \label{fig:VTSCalRes}
\end{figure}

To calibrate the added noise of the NKPA, we show the noise source and the detection chain in Fig. \ref{fig:VTSCalRes} (a). The intracavity photon occupancy can be represented as $n_{\textrm{mode}} = (\kappa_\mathrm{i}n_\mathrm{i} + \kappa_\mathrm{e}n_\mathrm{e})/\kappa$, incorporating both environmental noise and externally coupled noise. Here, $n_{\mathrm{i}}$ and $n_{\mathrm{e}}$ are dictated by the Bose-Einstein distribution $n_{\mathrm{s}}(T) = 1/\left(\exp(\hbar \omega_s/k_B T) - 1 \right)$, with corresponding temperatures being $T_\mathrm{dev}$ and $T_\mathrm{src}$. In the phase-preserving amplification mode, the output field is related to the input signal field $a_{\mathrm{S, mode}}$ and input idler field $a_{\mathrm{I, mode}}$ as  
\begin{equation}\label{equ:signal_idler}
        a_{\mathrm{out, S}} = \sqrt{G_\mathrm{N}} a_{\mathrm{S, mode}} + \sqrt{G_\mathrm{N} - 1}a_{\mathrm{I, mode}}^{\dag},
\end{equation}
where $G_\mathrm{N}$ is the NKPA gain. As we operate the NKPA in the high-gain limit, i.e., $G_\mathrm{N} \gg 1$, the added noise of the subsequent amplifiers such as HEMT and Room-temperature amplifier is negligible. With the assumption that the idler mode occupancy is the same as the signal mode, the output power spectral density is expressed as 
\begin{equation}
\begin{split}
        \frac{S_{\mathrm{out}}}{BW} & = \hbar \omega G (2n_{\mathrm{mode}} + 1), \\
                        & = \hbar \omega G\frac{\kappa_{\mathrm{e}}}{\kappa} (2n_{\mathrm{e}}(T_\mathrm{src})+ 2\frac{\kappa_{\mathrm{i}}}{\kappa_{\mathrm{e}}} n_{\mathrm{i}}(T_\mathrm{dev}) + \frac{\kappa}{\kappa_{\mathrm{e}}}),
    \label{equ:output_ref_e}
\end{split}
\end{equation}
where $S_{\mathrm{out}}$ is the power spectrum, $BW$ represent resolution bandwidth of the power spectrum, $G = G_\mathrm{R}G_\mathrm{H}G_\mathrm{N}$ is the total system gain, and $G_\mathrm{R}$ and $G_\mathrm{H}$ correspond to the gain of the room-temperature (RT) amplifier and HEMT. We fix the device physical temperature $T_\mathrm{dev} = 130$ mK, while varying the temperature of the VTS1 $T_\mathrm{src}$, during which we take measurements of the power spectrum. By fitting the normalized power spectrum in quanta ($S_{\mathrm{out}}\kappa/BW\cdot \hbar \omega G\kappa_\mathrm{e}$) as a function of VTS1 temperature $T_\mathrm{src}$, we are able to determine the added noise referred to the amplifier input: 
\begin{equation}
    n_{\mathrm{add}} = 2\frac{\kappa_{\mathrm{i}}}{\kappa_{\mathrm{e}}} n_{\mathrm{i}}(T_{\mathrm{dev}}) + \frac{\kappa}{\kappa_{\mathrm{e}}} - \frac{1}{2} + n_{\mathrm{exc}},
\end{equation}
where we subtract the $0.5$ quantum limit of noise quanta accompanying the input signal, and $n_{\mathrm{exc}}$ is the excess added noise not captured by the radiatively-cooled amplifier model. The above expression can be rewritten as
\begin{equation}
    n_{\mathrm{add}} = \frac{\kappa_{\mathrm{i}}}{\kappa_{\mathrm{e}}} (2n_{\mathrm{i}}(T_\mathrm{dev}) + 1) + 0.5 + n_{\mathrm{exc}}
    \label{equ:n_add_cal}
\end{equation}
The base temperature (130 mK) calibration is shown in Fig. \ref{fig:VTSCalRes} (b). Eq. (\ref{equ:n_add_cal}) suggests that with the same intrinsic and external coupling condition, the incremental added noise due to the increased device physical temperature can be expressed as 
\begin{equation}
        \Delta n_{\mathrm{add}} = 2\frac{\kappa_{\mathrm{i}}}{\kappa_{\mathrm{e}}} \left[ n_\mathrm{i}(T_{\mathrm{dev, high}}) - n_\mathrm{i}(T_{\mathrm{dev, low}}) \right].
\end{equation}
This equation shows that the elevated added noise is linearly dependent on the difference noise quanta and the internal/external coupling ratio. This contribution remains small with a very overcoupled device.

\subsection{Device performance at higher temperatures}
\begin{figure}[t]
    \centering
    \includegraphics[width=\linewidth]{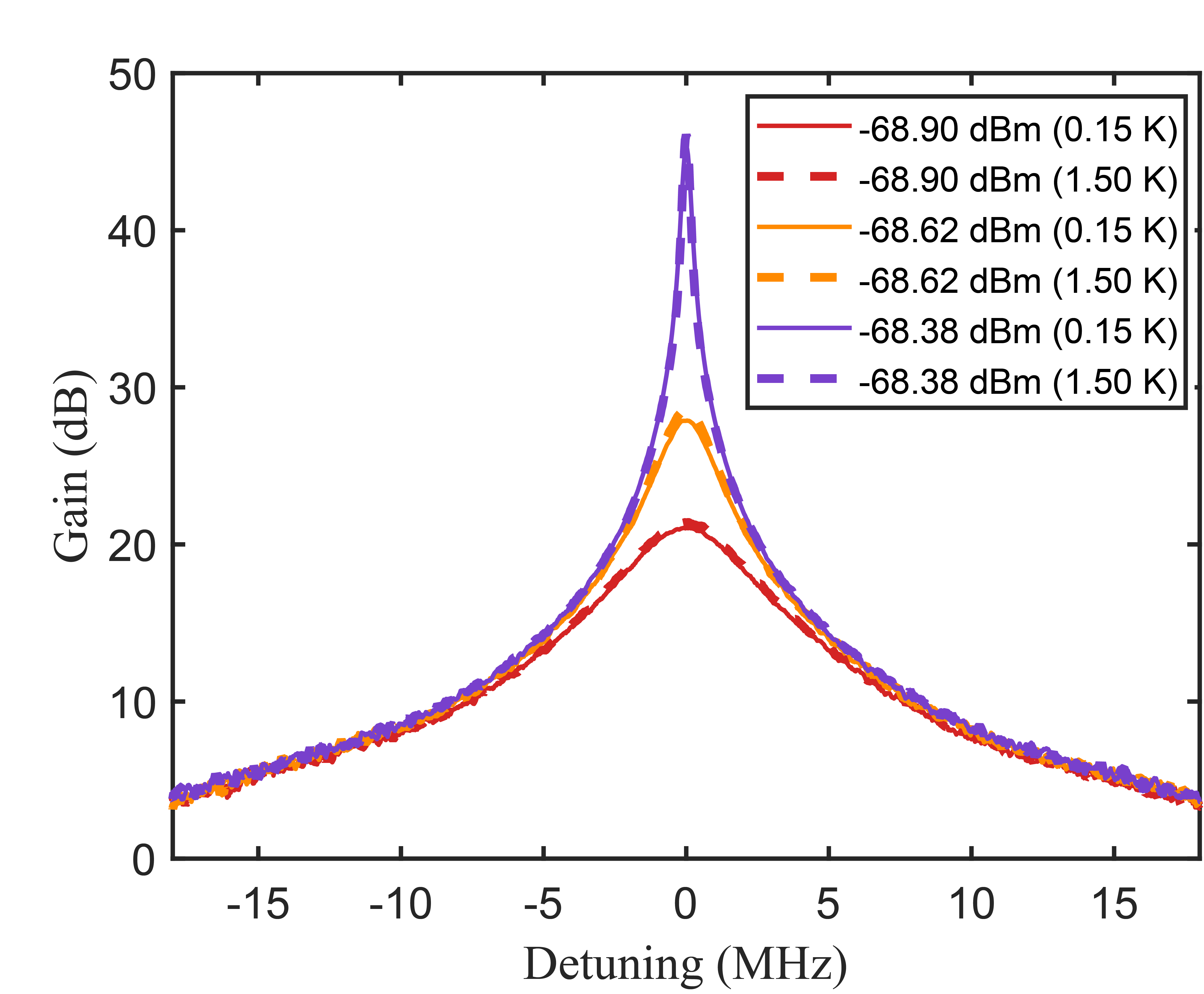}
    \caption{Amplifier gain characteristics at 150 mK (solid line) and 1.5 K (dashed line). Phase-preserving amplification as a function of the signal frequency detuning from the center frequency of the two pumps. More than 45 dB gain is observed with -68.90 dBm on-chip pump power at both temperatures. }
    \label{fig:gain_compare}
\end{figure}

\begin{figure}
    \centering
    \includegraphics[width=\linewidth]{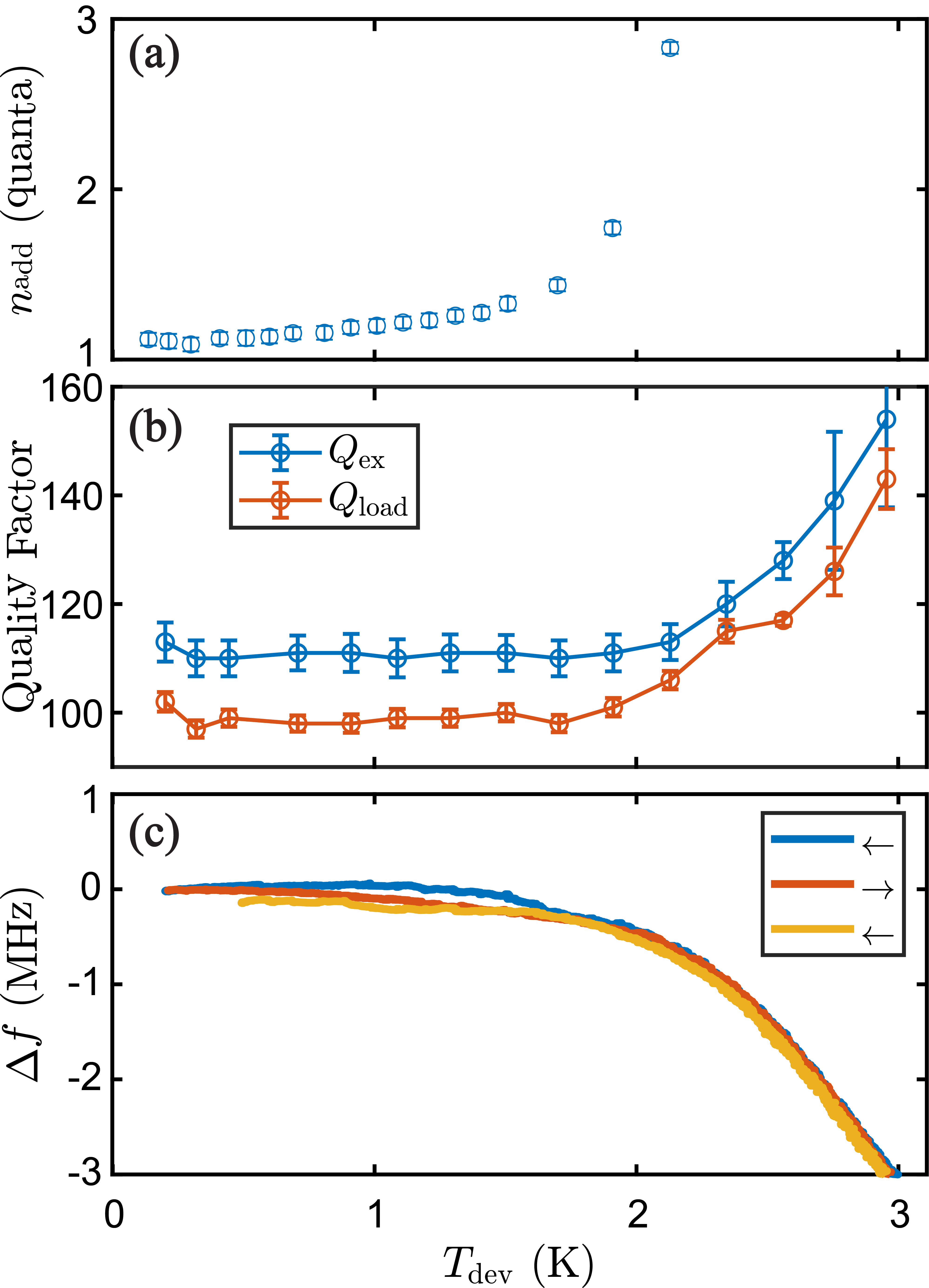}
    \caption{Device characteristics as a function of device physical temperature. (a) Added noise in quanta up to 2.12 K. (b) External Q and loaded Q extracted from the reflection spectrum fitting. (c) Phase-locked loop (PLL) tracking of frequency shift as a function of device physical temperature $T_\mathrm{dev}$ at 7.1509 GHz. The legend indicates the temperature sweeping direction. 
}
    \label{fig:PLL}
\end{figure}

In the main text, we demonstrate that the amplifier's added noise remains quantum-limited up to a physical temperature of 1.5 K. Impressively, the gain performance does not show any noticeable change, shown in Fig. \ref{fig:gain_compare}, where we plot the device gain as a function of detuning under various pump powers. When we compare the gain performance at 150 mK and 1.5 K, it becomes apparent from the figure that the gain profiles closely overlap under the same pump power. This result shows an impressive gain above 45 dB, achieved at both operating temperatures of 150 mK and 1.5 K.

The device performance, however, becomes less consistent for the temperature above 2 K, evident from the power spectrum result shown in Fig. \ref{fig:gainspectrum} in the main text, where a dramatic increase in the noise floor is observed. This NKPA performance degradation becomes even more evident with the added noise results from the NKPA, measured at operating temperatures up to 2.1 K, shown in Fig. \ref{fig:gain_compare}(a). The added noise rises steeply from 1.3 quanta at 1.5 K to more than 2.8 quanta at 2.1 K.

When the device temperature is increased to above 2 K, we also observe the deviation of the quality factor (Fig. \ref{fig:PLL} (b)) and resonance frequency shift (Fig. \ref{fig:PLL} (c)). Similar results have been reported in previous literature \cite{Zhou2014-cpl,PhysRevLett.121.117001}. Both the frequency shift and the change of internal Q could be attributed to the temperature-dependent dielectric constant due to TLSs in surface oxidation \cite{Zhou2014-cpl} or the effect of thermal quasiparticle \cite{PhysRevLett.121.117001}. The increase in the external Q is likely a result of the diminished capacitive coupling arising from elevated temperatures. 

In summary, the concurrent shift in added noise, quality factor, and frequency around 2 Kelvin indicates the emergence of spurious effects at elevated temperatures, due to either TLS or thermal quasiparticle. Higher $T_c$ superconductor materials such as NbTiN \cite{Yen1967-fb, malnou2022performance} or MBE-grown NbN \cite{Katzer2020-bs} can be explored to mitigate these effects, and potentially make near-quantum-limited amplifiers that work at even higher temperatures.

\subsection{System noise analysis for perspective applications}

\begin{figure}
    \centering
    \includegraphics[width=1\linewidth]{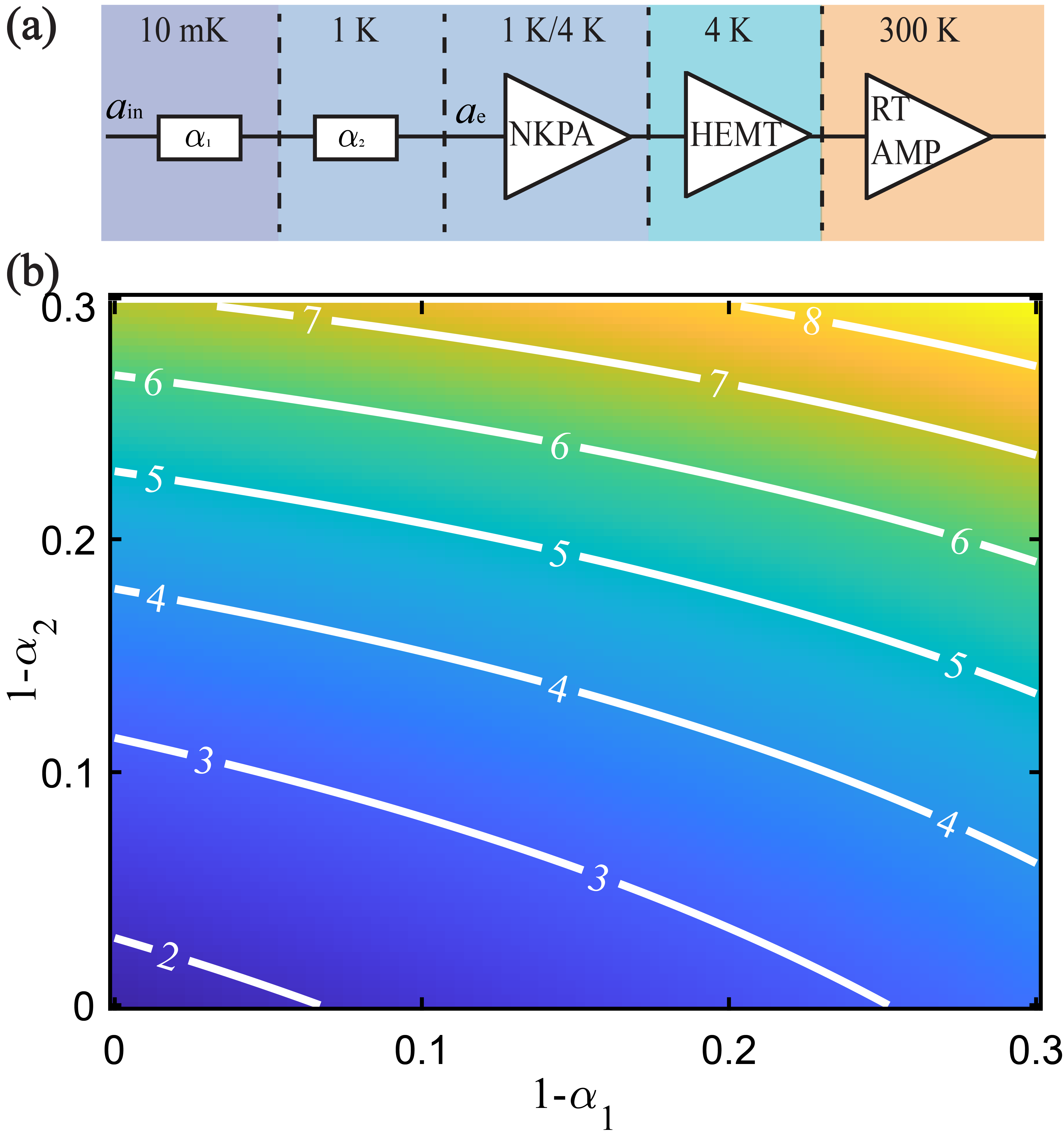}
    \caption{(a) The output line model for perspective application. The model includes 3 stages of amplifiers (NKPA, HEMT, and room-temperature (RT) amplifier), and imperfect transmission coefficients in the mixing chamber and 1-K plate before the amplifers, denoted as $\alpha_1$ and $\alpha_2$, respectively. (b) Color contour plot of the system added noise as a function of attenuations at mixing chamber and 1-K stage. The contour lines show the system added noise in quanta. }
    \label{fig:4kcontour}
\end{figure}
In perspective applications, where radiatively cooled NKPA is used for quantum signal readout, illustrated in Fig. \ref{application_diagram}, the additional insertion loss between the signal source (e.g. the qubit) and the first-stage amplifier could compromise the readout performance. Here we quantify the total system noise referred to the signal source at mixing chamber stage. The schematic diagram of the detection chain is shown in Fig. \ref{fig:4kcontour} (a). The input field $a_\mathrm{in}$ from MXC is transmitted to the NKPA installed at 1-K or 4-K plate, where the signal is being amplified. The signal then proceeds to a HEMT situated at the 4-K plate before reaching a room-temperature amplifier. We presume the NKPA's substantial gain can saturate the added noise from subsequent amplifiers. Therefore, we neglect added noise from the output line after NKPA.

We use $\alpha_1$ and $\alpha_2$ to represent the power transmission coefficients in the signal line at the MXC and the 1-K or 4-K stage respectively. In a practical minimal setup, $\alpha_1$ should include the loss of 4 connectors, and an isolator. $\alpha_2$ should include the loss of 6 connectors, a directional coupler, and a circulator. The attenuation thus couples the corresponding thermal fields $h_1$ and $h_2$ into the output. Then the input quanta to the NKPA is
\begin{equation}
    a_{\mathrm{e}} = \sqrt{\alpha_{2}} \left(\sqrt{\alpha_{1}}a_{\mathrm{in}} + \sqrt{1-\alpha_{1}}h_{1}^\dagger \right) + \sqrt{1 - \alpha_{2}}h_{2}^\dagger.
    \label{equ:mode_in_to_e}
\end{equation}
where $a_{\mathrm{in}}$ is the input signal at MXC. The equilibrium noise photon occupation $\langle h_{\mathrm{1}}^\dagger h_{\mathrm{1}}\rangle = n_\mathrm{s}\mathrm{(0.01K)}$ and $\langle h_{\mathrm{2}}^\dagger h_{\mathrm{2}}\rangle = n_\mathrm{s}\mathrm{(1K)}$. The NKPA device could be mounted on either the 1-K plate or the backside of the 4-K plate, as needed by the perspective applications. Therefore, the intra-cavity photon number of the NKPA can be expressed as 
\begin{equation}
    n_\mathrm{mode}= \frac{\kappa_{\mathrm{e}}}{\kappa} \langle a_{\mathrm{e}}^\dagger a_{\mathrm{e}} \rangle + \frac{\kappa_{\mathrm{i}}}{\kappa} n_\mathrm{s}(T_\mathrm{dev})
    \label{equ:intra_cavityField}
\end{equation}
Plugin Eq. (\ref{equ:intra_cavityField}) and Eq. (\ref{equ:mode_in_to_e}) into Eq. (\ref{equ:output_ref_e}) for both signal and idler modes, we can derive the system noise as
\begin{equation}
\begin{split}
     n_{\mathrm{sys}} = & \frac{1}{\alpha_1 \alpha_2} \left[ \frac{\kappa_\mathrm{i}}{\kappa_\mathrm{e}} \left(2n_\mathrm{s}(T_\mathrm{dev})+\frac{1}{2}\right) + n_{\mathrm{exc}} + 1\right] \\
     + & \frac{\kappa}{\kappa_e} \left[ 2\frac{1 - \alpha_1}{\alpha_1} \left( n_\mathrm{s}(0.01K) +1\right)  + 2\frac{1 - \alpha_2}{\alpha_1 \alpha_2} \left( n_\mathrm{s}(1K) + 1\right) \right]
\end{split}
\label{equ:final_n_add}
\end{equation}
The terms inside the first square bracket represent the amplifier noise, with a prefactor of $1/\alpha_1\alpha_2$ meaning it is effectively amplified due to attenuated signal. The two terms inside the second square bracket are the coupled environment noise from MXC and 1-K plate due to loss. With $\alpha_1 \rightarrow 1$ and $\alpha_2 \rightarrow 1$, Eq. \ref{equ:final_n_add} reduce to Eq. \ref{equ:n_add_cal} plus 0.5 to account for the vacuum noise accompanied by the signal. 

In figure 8b, we illustrate the system noise as a function of attenuations in MXC and 1-K plate ($1-\alpha_1$ and $1-\alpha_2$), while maintaining a fixed ration between coupling rates $\kappa_\mathrm{e}/\kappa = 0.98$. The dense contour line along the y-axis implies a higher sensitivity of system noise to the loss at the 1-K plate. The system noise remains under 6 quanta if the losses at MXC and 1-K are each below 1 dB ($1-\alpha_1<0.1$, $1-\alpha_2<0.1$). If both attenuations reach 1.5 dB, the system noise would exceed 8 quanta. 

\bibliographystyle{apsrev4-1}
\bibliography{References.bib}

\end{document}